\documentclass[prl,twocolumn,superscriptaddress]{revtex4-2}

\usepackage{graphicx}
\usepackage{dcolumn}
\usepackage{bm}
\usepackage{subfigure}
\usepackage{float}
\usepackage{booktabs}
\usepackage{siunitx}
\usepackage[colorlinks=true,linkcolor=blue,citecolor=blue, urlcolor=blue]{hyperref}
\usepackage{upgreek}

\usepackage{xcolor}
\begin{document}

\preprint{APS/123-QED}

\title{Haloscope Searching for Dark Photons at Q-band with a Novel Coupling Tuning Structure}

\author{Yue Yin}
\affiliation{Laboratory of Spin Magnetic Resonance, School of Physical Sciences, University of Science and Technology of China, Hefei 230026, China}
\affiliation{Anhui Province Key Laboratory of Scientific Instrument Development and Application, University of Science and Technology of China, Hefei 230026, China}
\affiliation{Hefei National Laboratory, University of Science and Technology of China, Hefei 230088, China}
\author{Runqi Kang}
\affiliation{Laboratory of Spin Magnetic Resonance, School of Physical Sciences, University of Science and Technology of China, Hefei 230026, China}
\affiliation{Anhui Province Key Laboratory of Scientific Instrument Development and Application, University of Science and Technology of China, Hefei 230026, China}
\affiliation{Hefei National Laboratory, University of Science and Technology of China, Hefei 230088, China}

\author{Man Jiao}
\email{man.jiao@zju.edu.cn}
\affiliation{Institute for Advanced Study in Physics, Zhejiang University, Hangzhou 310027, China}
\affiliation{Institute of Quantum Sensing and School of Physics, Zhejiang University, Hangzhou 310027, China}

\author{Xing Rong}%
\email{xrong@ustc.edu.cn}
\affiliation{Laboratory of Spin Magnetic Resonance, School of Physical Sciences, University of Science and Technology of China, Hefei 230026, China}
\affiliation{Anhui Province Key Laboratory of Scientific Instrument Development and Application, University of Science and Technology of China, Hefei 230026, China}
\affiliation{Hefei National Laboratory, University of Science and Technology of China, Hefei 230088, China}
\affiliation{Zhejiang Key Laboratory of R\&D and Application of Cutting-edge Scientific Instruments, Zhejiang University, Hangzhou, 310027, China}


\begin{abstract}
Laboratory searching for dark matter is crucial for understanding several fundamental conundrums in physics and cosmology.
Most cavity-based haloscope searches focus on the frequency range below 10 GHz, while the parameter space with higher frequency remains rarely explored, due to the challenges lying in the fabrication of microwave cavities.
Here we report the first Q-band haloscope searching for dark photons with a 33.141 GHz cavity.
A novel coupling tuning structure separated from the cavity was designed so as not to degrade the quality factor of the cavity.
We have established the most stringent constraints $\chi<2.5\times10^{-12}$ at a confidence level of 90 $\%$ in the frequency range from 33.139 GHz to 33.143 GHz, corresponding to the mass of dark photons ranging from 137.05 $\upmu$eV to 137.07 $\upmu$eV. The results surpass the previous astronomical constraints by nearly three orders of magnitude. This work has demonstrated the feasibility of dark matter haloscopes at Q band. In the future, the constraints can be further improved by more than one order of magnitude through low-temperature experiments, and the setup can be extended to search for axions, axion-like particles, and high-frequency gravitational waves.

\end{abstract}

\maketitle

Many astronomical observations indicate the existence of dark matter\cite{tucker19981e,clowe2004weak}. It is widely accepted that dark matter makes up 24\% of our universe, much more than ordinary matter. Dark matter is also believed to play an important role in the early evolution of the universe\cite{kuster2007axions}. However, despite the indirect evidence from the gravitational effects\cite{tucker19981e,clowe2004weak}, the nature and the properties of dark matter remain a mystery. Many candidates for dark matter have been proposed. Among them the dark photon is one of the most attracting candidates since it is suggested by many theories beyond the Standard Model, and can help to explain many experimental and astronomical anomalies, including the velocity discrepancy of galaxies\cite{zwicky1933redshift} and the cosmic ray anomaly\cite{chang2008excess,cholis2009high}. Muon anomalous magnetic moment\cite{abi2021measurement,cazzaniga2021probing} and W-boson mass anomaly\cite{cdf2022high,thomas2022constraints} can also be explained by the existence of dark photons.

The dark photon is a spin-1 boson which derives from a new dark U(1) added into the standard model gauge group\cite{galison1984s,holdom1986two,holdom1986searching}. Its coupling to the ordinary photon can be described by the Lagrangian:
\begin{eqnarray}
    \mathcal{L}=-\frac{1}{4}F_{\mu\nu}^2-\frac{1}{4}V_{\mu\nu}^2-\frac{m_{A'}}{2}^2A_{\mu}'A'^{\mu}+\frac{\chi}{2}F_{\mu\nu}V^{\mu\nu},
    \label{eq:L}
\end{eqnarray}
where $A_{\mu}$ and $A_{\mu}'$ are the gauge fields of ordinary photons and dark photons, $F_{\mu\nu} = \partial_{\mu}A_{\nu}-\partial_{\nu}A_{\mu}$ and $V_{\mu\nu}=\partial_{\mu}A_{\nu}'-\partial_{\nu}A_{\mu}'$ are the corresponding field tensors, $m_{A'}$ is the mass of dark photons, and $\chi$ is a dimensionless parameter describing the kinetic mixing between ordinary photons and dark photons. Both $\chi$ and $m_{A'}$ are free parameters.
Previous calculations suggest that there exists a promising dark photon mass range from several $\upmu$eV to several hundred µeV\cite{borsanyi2016calculation,kawasaki2015axion}, while the value of $\chi$ is believed to be very small. Therefore, searching for dark photons is extremely challenging.

The haloscope is one of the most popular methods for dark photon searching\cite{sikivie1983experimental,sikivie1985detection,caputo2021dark}. It is based on the assumption that the Milky Way is immersed in a dark matter halo with the density $\rho_{A'}=0.45\ {\rm GeV/cm^3}$\cite{read2014local}. By manually breaking the translation invariance through metallic or dielectric instruments, dark photons can be continuously converted into ordinary photons and then detected. For a metallic cavity, the output power induced by the dark photon field can be written as:
\begin{eqnarray}
    P_s(\nu)=2\pi\nu_{A'}\rho_{A'}\chi^2VC\frac{Q_{\rm L}Q_{\rm a}}{Q_{\rm L}+Q_{\rm a}}\frac{\beta}{1+\beta}L(\nu,\nu_{\rm c},Q_{\rm L}),
    \label{eq:power}
\end{eqnarray}
where $\nu_{A'}=m_{A'}/h$ is the frequency of the dark photon field, $h$ is the Planck's constant, $V$ is the volume and the form factor of the cavity, and $C=\langle\cos^2\theta\rangle|\int dV\boldsymbol{E}|^2/V\int dV E^2$ is the form factor, where the $\theta$ is the angle between the polarization direction of the dark-photon field and the electric field. $Q_{\rm a}=10^6$ is the quality factor of the dark photon field. $Q_{\rm L} = Q_0/(1+\beta)$ is the loaded factor of the cavity, where $Q_0$ is the intrinsic quality factor and $\beta$ is the coupling strength between the cavity and the output port. $L(\nu,\nu_{\rm c},Q_{\rm L})=\left[1+4Q_L^2(\nu/\nu_{\rm c}-1)^2\right]^{-1}$ is the Lorentzian line shape, where $\nu_{\rm c}$ is the resonant frequency of the cavity.
\begin{figure}[htp]
    \centering
    \includegraphics[width=0.9\linewidth]{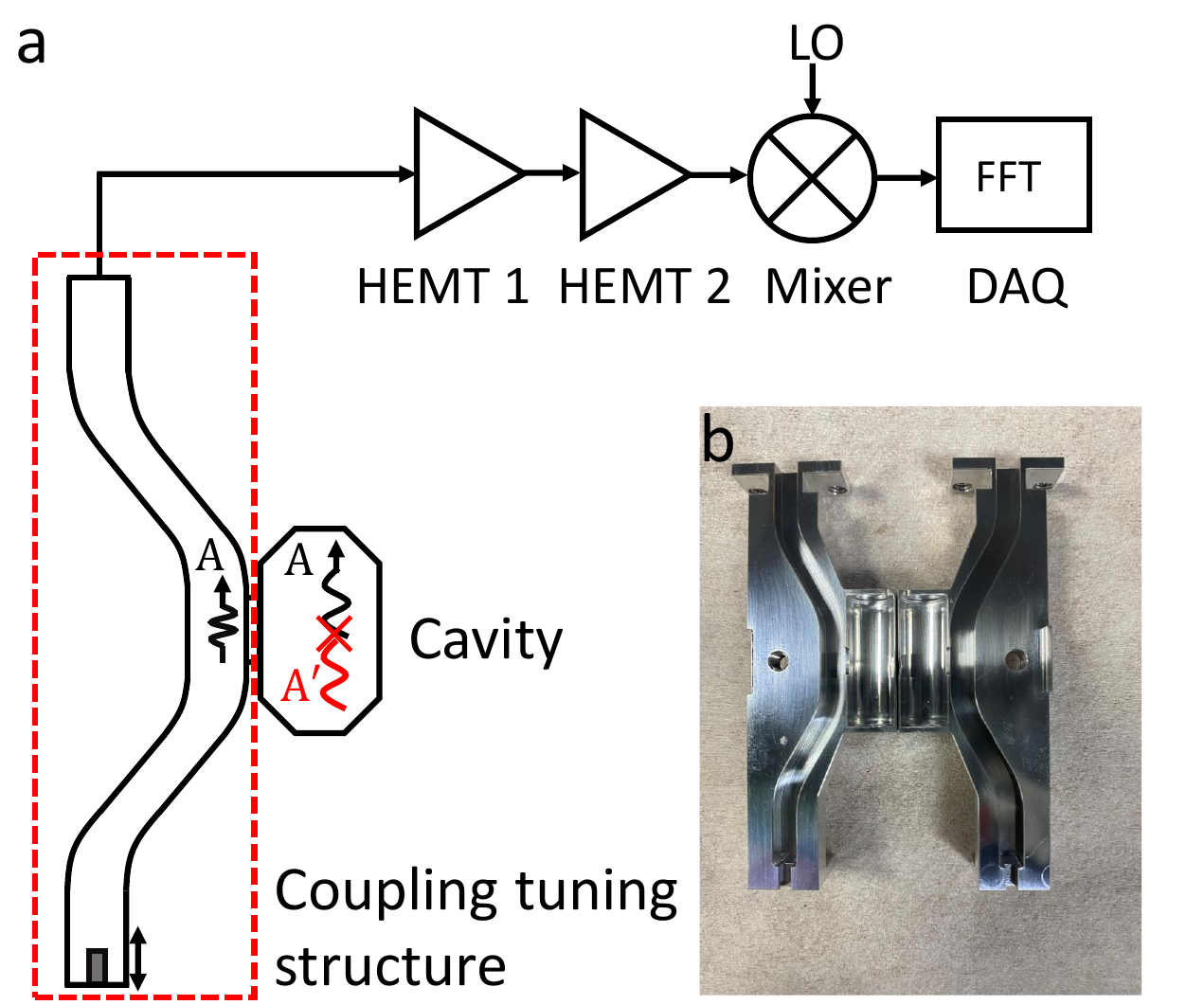}
    \caption{Schematic diagram of the experimental setup. The setup consisted of a cylindrical cavity, a bow-shaped waveguide, a cascade of HEMTs, a mixer and a DAQ system. At the end of the waveguide was a movable metallic brubaker2017firsbulk used to adjust the coupling strength between the cavity and the waveguide.}
    \label{fig:setup}
\end{figure}
Resonant detections are conducted by groups such as ORGAN\cite{quiskamp2024exclusion}, QUAX\cite{alesini2021search,alesini2022search}, CAPP\cite{kim2024experimental}, HAYSTAC\cite{bai2024dark}, ADMX\cite{bartram2021search,cervantes2022search} and SHANHE\cite{tang2024first} with microwave cavities. Experiments like FUNK\cite{andrianavalomahefa2020limits}, BREAD\cite{liu2022broadband}, DOSUE\cite{kotaka2023search}, QUALIPHIDE\cite{ramanathan2023wideband}, BRASS-p\cite{bajjali2023first}, atomic magnetometers\cite{jiang2024long} and Dark SRF\cite{romanenko2023search} achieved large-width searching for dark photon through non-resonant devices. These projects focus on the frequency range below 10 GHz, corresponding to the dark photon mass range below 40 $\upmu$eV. However, dark photons with the mass larger than 40 $\upmu$eV are rarely explored, despite their strong motivation. The prime challenge to searching for dark photons at high frequency comes from the frequency-elength relationship. As the Compton wavelength of dark photons is inverse to the frequency, the effective volume of the haloscope cavities shrinks with the resonant frequency $\nu_{\rm c}$ as $\nu_{\rm c}^{-3}$. As a result, at high frequencies, the dark-photon-induced signal would be very weak and the fabrication of the cavities will be difficult. Recently, the MADMAX group introduced dielectric cavities to search for dark photons around 19 GHz\cite{garcia2024first}. The ORGAN group used high-frequency metallic cavities to search for dark photons as well as axions in the frequency range from 26 GHz to 27 GHz\cite{quiskamp2024exclusion}. Meanwhile, research on the frequency range above 30 GHz remains insufficient.

In this paper, we report the first haloscope experiment at Q-band.
A metallic cavity with the resonant frequency of 33.141 GHz was employed, corresponding to the dark photon mass of $137.06$ $\upmu$eV. A novel coupling tuning structure separated from the cavity was designed, in order to realize coupling tuning without degrading the quality factor of the cavity.
This work sets the most stringent constraints on dark photons at the $2.5\times10^{-12}$ level in a mass range of 16.5 neV around 137.06 $\upmu$eV, exceeding the previous results by nearly three orders of magnitude.

Figure \ref{fig:setup} shows the diagram of the experimental setup. It was housed at room temperature. The setup consisted of a cylindrical cavity, a transmission waveguide, a cascade of high-electron-mobility transistor (HEMT) amplifiers, a mixer and a data acquisition system (DAQ). The the cavity was cylindrical with diameter 3.56 mm and length 20.00 mm. Since the Lagrangian in Eq. \ref{eq:L} allows for the spontaneous conversion between dark photons and ordinary photons, the cavity could collect the microwave photons converted from dark photons and feed them into the readout circuit. Only the cavity modes with a non-zero form factor can be excited by dark photon fields. In this work the ${\rm TM_{010}}$ mode with a form factor of 0.23 was utilized.

A coupling tuning structure is necessary because according to Eq. \ref{eq:power} the signal power is maximized at critical coupling, i.e., $\beta = 1$. Since the cavity was small and the fabrication was challenging, any coupling tuning structure inside the cavity would greatly degrade its quality factor. Therefore, a bow-shaped waveguide was designed to couple the dark-photon-induced microwave power into the readout chain. The waveguide cross-section was a 7.1 mm $\times$ 3.6 mm rectangle. At the end of the waveguide was a movable metallic bulk. As the position of the metallic bulk could decide the positions of the nodes and antinodes of the microwave in the waveguide, the coupling strength between the cavity and the waveguide could be adjusted by moving the metallic bulk.
The dark-photon-induced microwave signal would be fed into the readout chain at the other end of the waveguide through a waveguide-to-coax adapter. The signal was first amplified by the HEMTs, then down-converted to around 2 MHz, and finally collected by the DAQ. The DAQ could perform the fast Fourier transformation and save the frequency spectra with an 100\% duty cycle\cite{kang2024near,tong2020high}.

\begin{figure}[t]
    \centering
    \subfigure{\includegraphics[width = 0.49\linewidth]{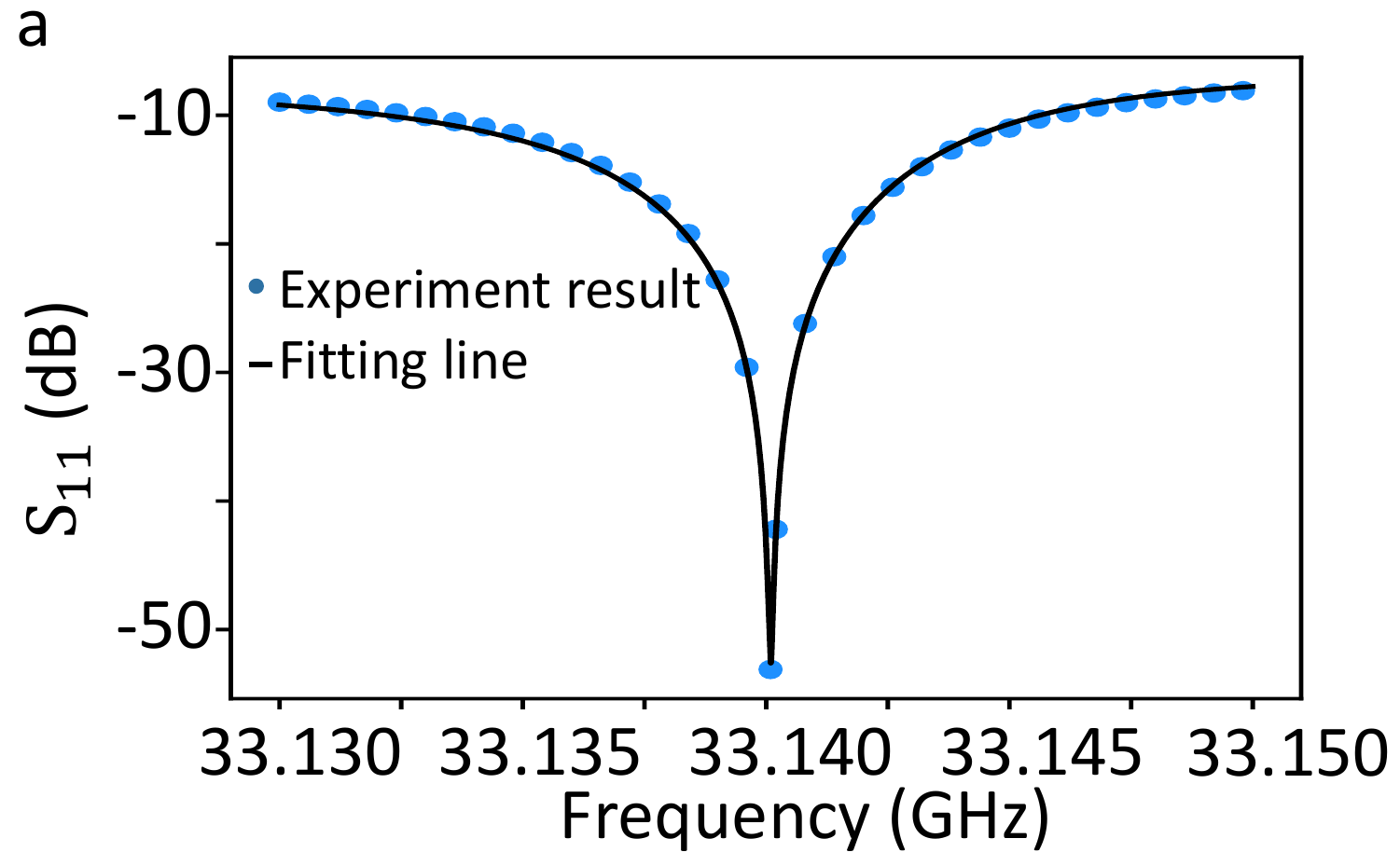}}
    \subfigure{\includegraphics[width = 0.49\linewidth]{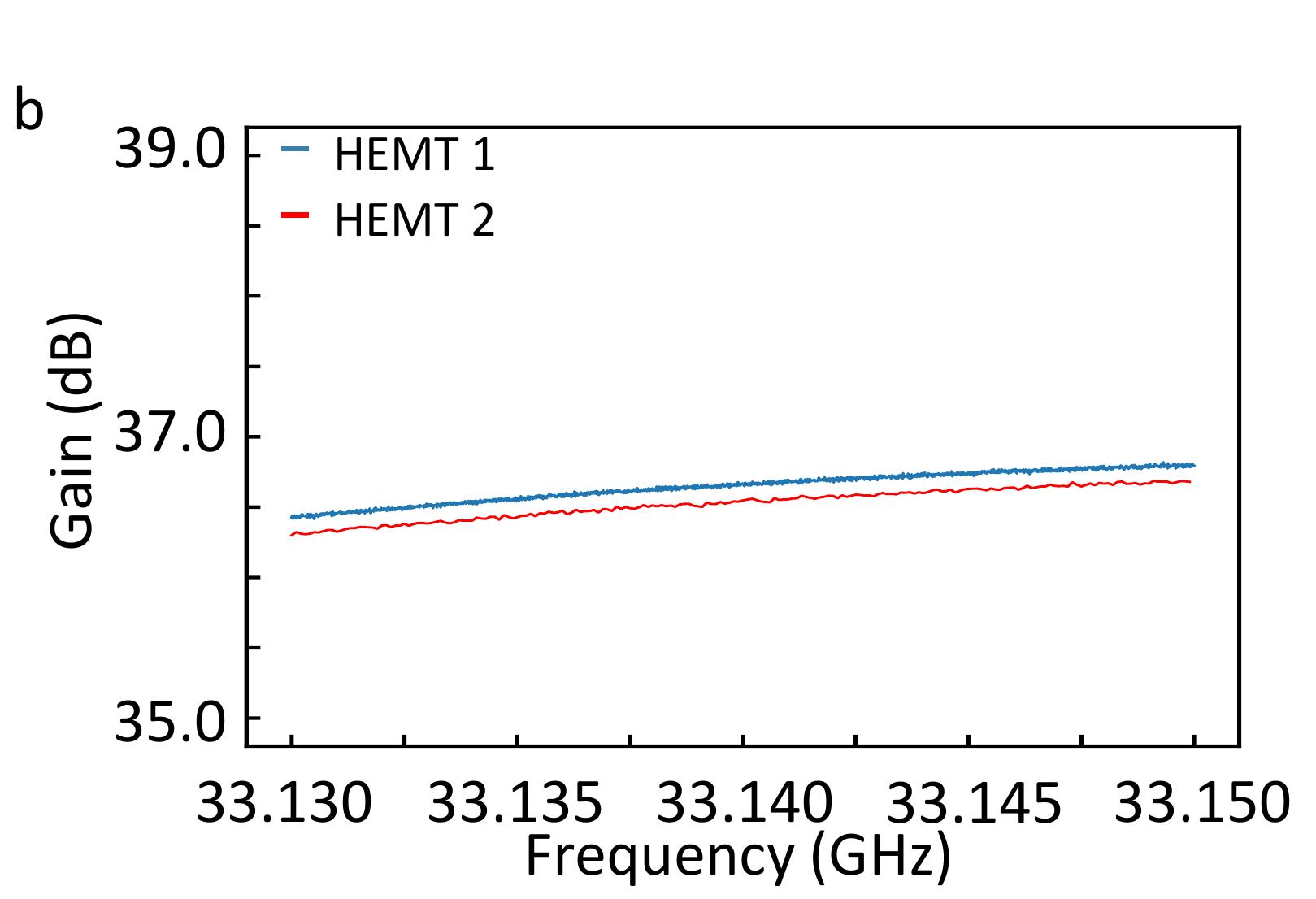}}
    \caption{Calibration results. (a) Reflection coefficients of the cavity. The blue dots are the experimental results and the black line is the fitting result. (b)The gain curves of the first HEMT (blue) and the second HEMT (red).}
    \label{fig:exp}
\end{figure}

In order to determine the resonant frequency $\nu_{\rm c}$, the loaded quality factor $Q_{\rm L}$, and the coupling strength $\beta$ of the cavity, the reflection coefficient of the output port $S_{11}$ was measured, as shown in Fig. \ref{fig:exp}(a). The result is fitted to the equation\cite{kudra2020high}:
\begin{eqnarray}
    S_{11}=20\log_{10}\left|\alpha\left(1-\frac{2\beta/(1+\beta e^{i\phi})}{1+2iQ_{\rm L}(\nu/\nu_{\rm c}-1)}\right)\right|,
\end{eqnarray}
where $\alpha$ is the power loss during transmission, and $\phi$ describes the impedance mismatch, which makes the spectrum asymmetric. By moving the metallic bulk we approached critical coupling with $\beta = 1.0243\pm 0.0007$. The resonant frequency and the quality factor of the cavity were $\nu_{\rm c}=3.3141\times10^{1}\pm 1.1228\times10^{-5} \;{\rm GHz}$ and $Q_{\rm L}=2520\pm 5$, respectively.

The noise power can be described by the equation:
\begin{eqnarray}
    P_{\rm t} = G_1G_2P_{\rm n,1}+G_2P_{\rm n,2}+P_{\rm n,3},
\end{eqnarray}
where $P_{\rm t}$ is the total noise power, $P_{\rm n,1}, P_{\rm n,2}, P_{\rm n,3}$ are the noise power of the first HEMT, the second HEMT and the DAQ system, respectively, and $G_{1,2}$ refer to the gains of the first and the second HEMT, respectively. Usually, the gain of HEMTs is large enough so that the total noise is dominated by the noise of the first HEMT. The gain curves of HEMT 1 (blue line) and HEMT 2 (red line) are shown in Fig. \ref{fig:exp}(b). In the frequency range from 33.130 GHz to 33.150 GHz, the gains of HEMT 1 and 2 are in the range of 36.42 dB$\sim$36.79 dB and 36.29 dB$\sim$36.68 dB, respectively.

\begin{figure}[htp]
    \centering
    \subfigure{\includegraphics[width = 0.49\linewidth]{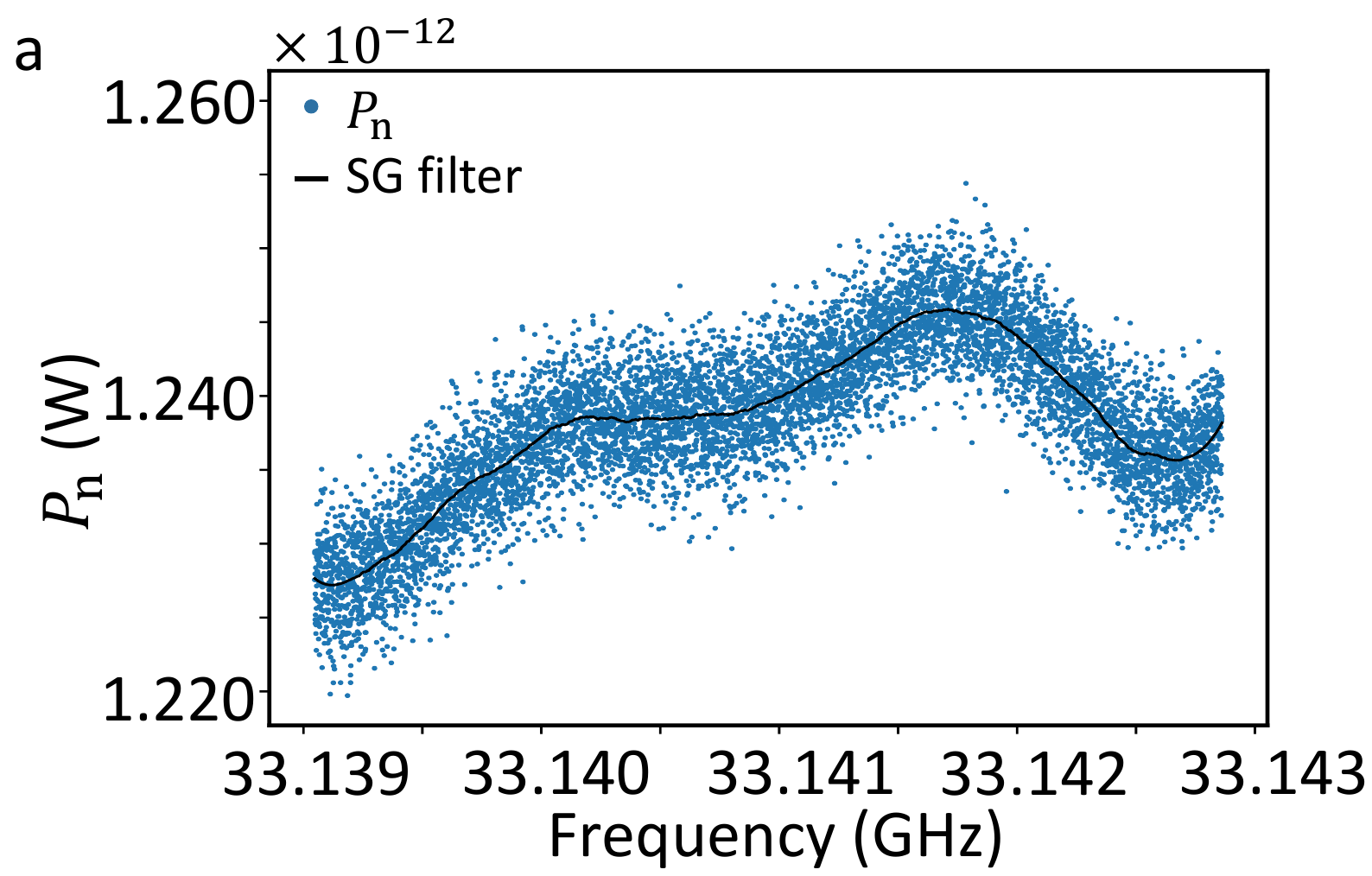}}
    \subfigure{\includegraphics[width = 0.49\linewidth]{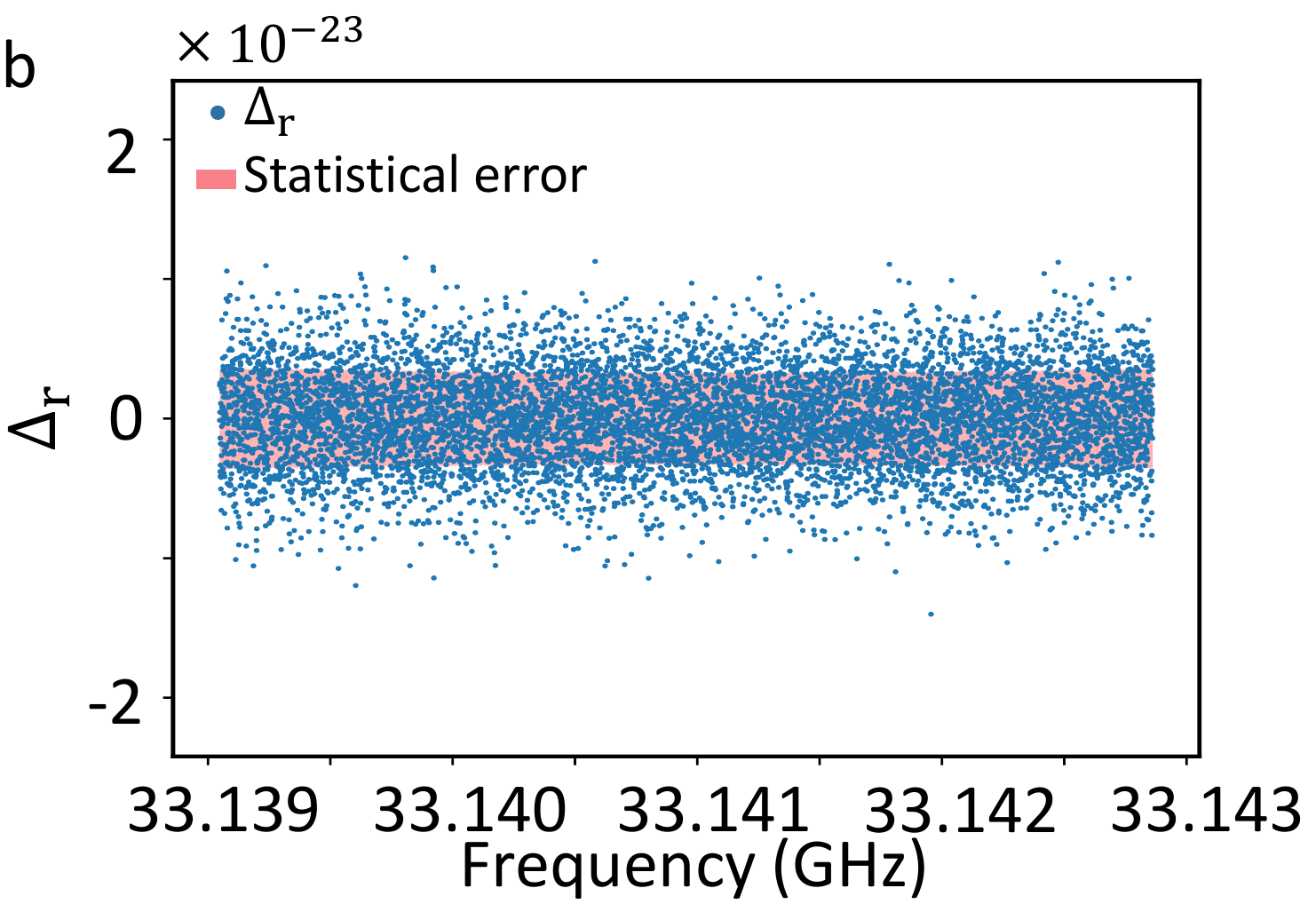}}
    \subfigure{\includegraphics[width = 0.49\linewidth]{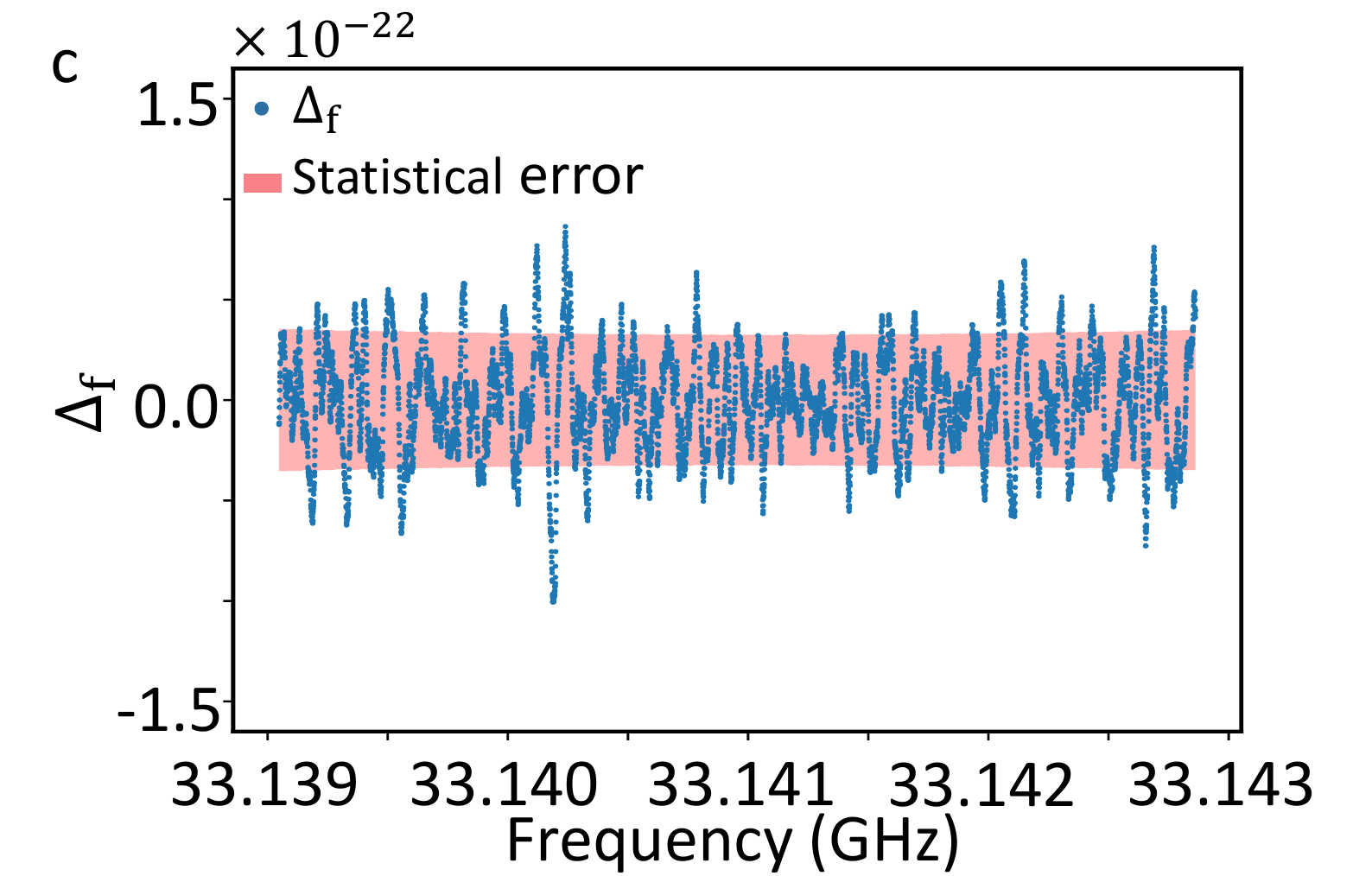}}
    \subfigure{\includegraphics[width = 0.49\linewidth]{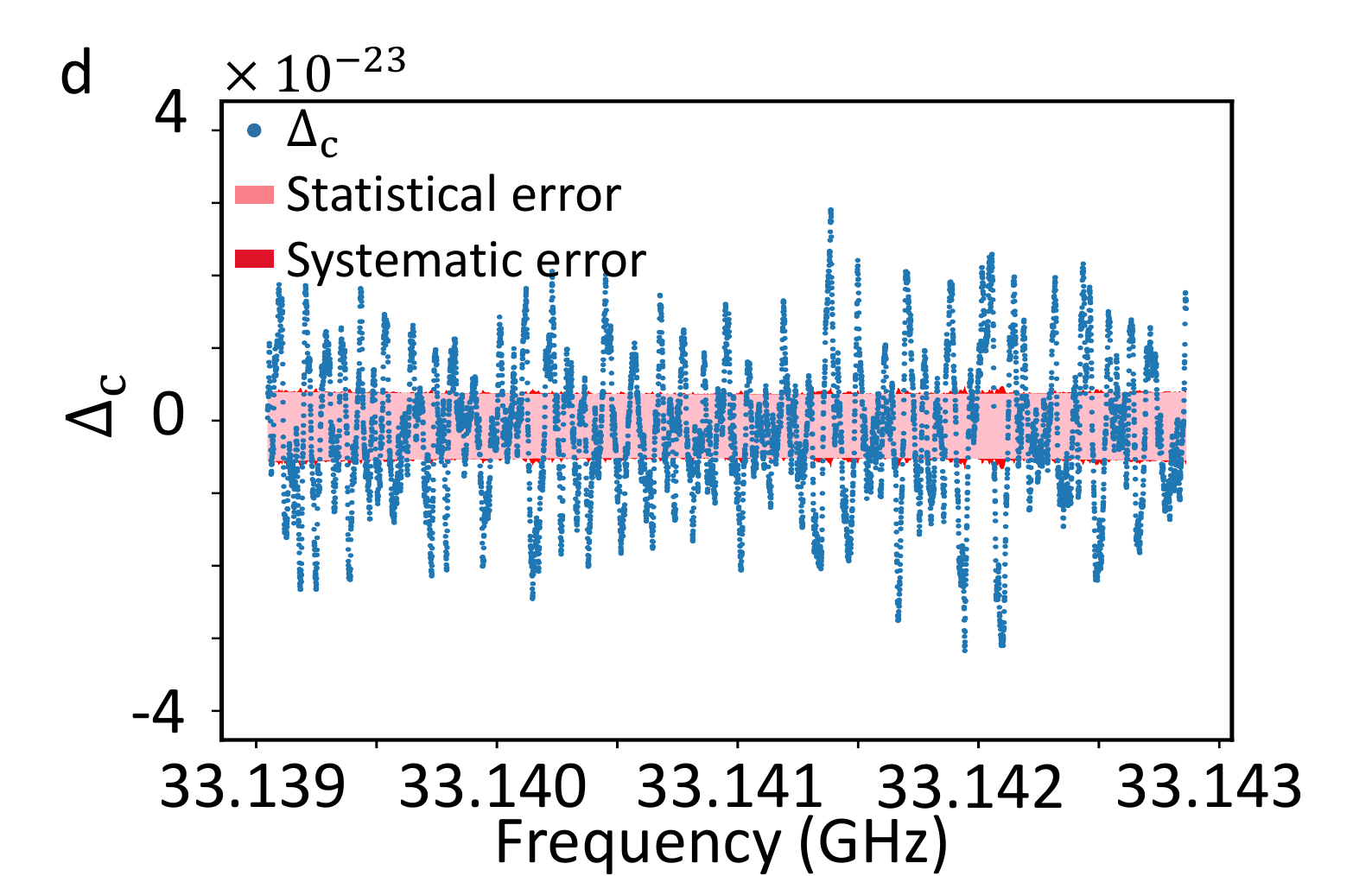}}
    \caption{Experiment results. (a) The noise power spectrum (blue dots) and the SG filter (black line) of the first subset. (b) The rescaled power excess (blue dots) and the statistical error (pink ribbon) of the first subset. (c) The convolved power excess (blue dots) and the statistical error (pink ribbon) of the first subset. (d) The averaged power excess (blue dots), the statistical error (pink ribbon), and the systematic errors (red ribbon) obtained from all subsets.}
    \label{fig:result}
\end{figure}

The dark photon searching experiment lasted for 12 hours, resulting in twenty thousand raw spectra. Although the signal-to-noise ratio would be optimized when the frequency bin width $B$ equals the linewidth of the dark photon field $\Delta\nu=33.141\; {\rm kHz}$, during the experiment $B$ was set to 477 Hz, so that a dark photon signal would consist of around 70 points and could be further verified through its line shape. The raw spectra underwent a coarse process with each neighbouring 1000 spectra divided into a subset and averaged into a single power spectrum. The blue dots in Fig. \ref{fig:result}(a) show the averaged power spectra $P_{\rm n}$ of the first subset. The curved baseline is due to the variance in the transmission efficiency of the readout line. Since only the power excess instead of the noise power itself is concerned, the SG filter $P^{\rm SG}$ was applied to each averaged spectrum\cite{PhysRevD.96.123008}, as the black line in Fig. \ref{fig:result}(a) shows. Since the power of a dark photon signal depends on its central frequency, the power excess was rescaled to the expected dark photon signal with $\chi=1$ through the equation:
\begin{equation}
    \Delta_{\rm r} = \frac{P_{\rm n}-P^{\rm SG}}{P_{\rm DP}(\nu,\chi=1)},
\end{equation}
where $\Delta_{\rm r}$ is the rescaled power excess, $P_{\rm DP}(\nu,\chi=1)$ is dark photon signal with $\chi=1$. The rescaled power excess and the corresponding standard deviation of the first subset are shown as the blue dots and the pink ribbon in Fig. \ref{fig:result}(b), respectively. Herein we set the threshold for a dark photon candidate signal to be 5 times of the standard deviation. As no normalized power excess over 5 times of standard deviation is observed, an exclusion of the kinetic mixing between dark photons and ordinary photons can be obtained from our results.
Since the frequency bin width of the spectra was only 1/70 of the linewidth of dark photons, the signal-to-noise ratio would be degraded by a factor of $\sqrt{70}$. In order to recover the signal-to-noise ratio, a convolution integration with the kernel being the normalized line shape of dark photons is applied to the rescaled power excess spectra\cite{cervantes2022admx}. Figure \ref{fig:result}(c) shows the convolved power excess of the first subset.  As Fig. \ref{fig:result}(c) shows, the blue dots indicate the convolved power excess $\Delta_{\rm f}$ and the pink ribbon represents the standard deviation $\sigma_{\rm f}$.

All convolved spectra have been averaged again to further suppress the noises, as shown in Fig. \ref{fig:result}(d). The averaged power excess $\Delta_{\rm c}$ is plotted in blue dots, and the statistical errors $\sigma_{\rm c}$ is drawn in pink ribbon. Here, systematic errors are introduced into the analysis. The mean values and relative uncertainties of the experimental parameters are listed in Table \ref{tab:parameters}. The statistical error and the systematic errors are combined as $\sigma_c' = \sqrt{\sigma_{\rm c}^2+\Delta_{\rm c}^2(\sigma_{Q_{\rm L}}^2+\sigma_{\beta}^2+\sigma_{\nu_{\rm c}}^2+\sigma_{V}^2)}$, where $\sigma_{Q_{\rm L}}, \sigma_{\beta}, \sigma_{\nu_{\rm c}},\sigma_{V}$ are the relative uncertainties of $Q_{\rm L}, \beta, \nu_{\rm c}, V$, respectively. The contribution of systematic errors is shown as the red ribbon in Fig. \ref{fig:result}(d).

\begin{table}[h!]
\centering
\begin{tabular}{l c S[table-format=1.1e2]}
\toprule
\text{Parameter} & \text{Value} & \text{Relative Uncertainty} \\
\midrule
$Q_{\rm L}$ & $2.5\times 10^{3}$ & 1.9e-3 \\
$\beta$ & 1.0 & 6.8e-4 \\
$\nu_{\rm c}$ & \SI{33.1}{\giga\hertz} & 3.6e-7 \\
$V$ & \SI{7.3e-4}{\liter} & 6.4e-2 \\
\bottomrule
\end{tabular}
\caption{Parameters and their values with relative uncertainties.}
\label{tab:parameters}
\end{table}

\begin{figure}[t]
    \centering
    \includegraphics[width=\linewidth]{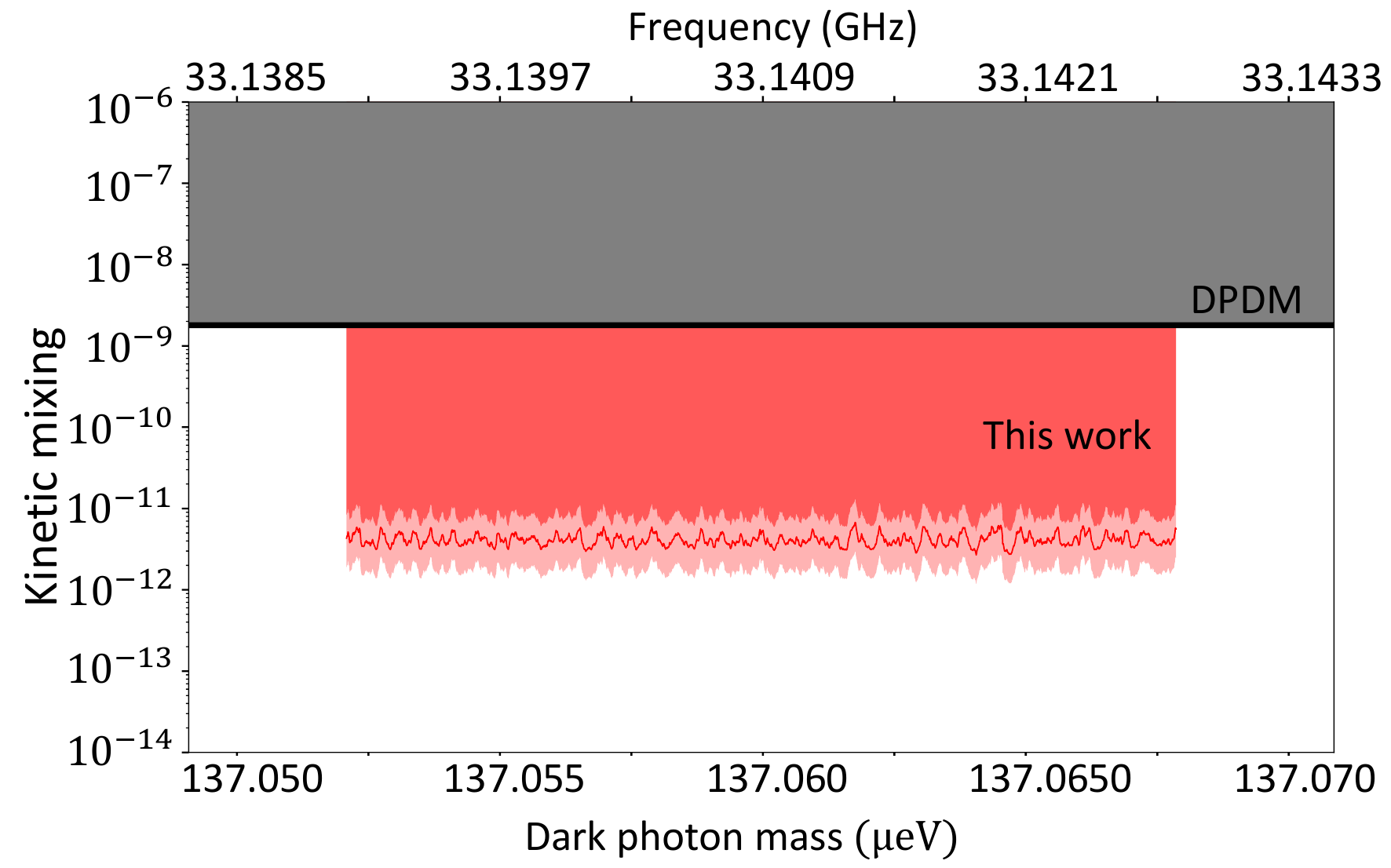}
    \caption{Constraints on the kinetic mixing between dark photons and ordinary photons. The red line refers to the upper limit of the kinetic mixing in the random polarization scenario. The shallow and dark red area refer to the excluded parameter spaces in the linear polarization scenario in the best and the worst condition, respectively. The gray area refer to the result DPDM\cite{Arias2012WISPyCD}.}
    \label{fig:limit}
\end{figure}

The constraints on the kinetic mixing $\chi$ can be calculated according to $\Delta_{\rm c}$ and $\sigma_{\rm c}$ obtained in Fig. \ref{fig:result}(d) by solving the equation:
\begin{equation}
    \int_0^{\chi^2_{90\%}}p(\chi^2|\Delta_{\rm c}){\rm d}\chi^2=90\%,
\end{equation}
where $\chi_{90\%}$ is the constraints on the kinetic mixing with a confidence level of 90$\%$, and the conditioned distribution of $\chi^2$ is
\begin{equation}
    p(\chi^2|\Delta_{\rm c})=\frac{\exp\left(-\frac{(\Delta_{\rm c}-\chi^2)^2}{2\sigma_{\rm c}^2}\right)}{\int_0^{\infty}\exp\left(-\frac{(\Delta_{\rm c}-\chi^2)^2}{2\sigma_{\rm c}^2}\right){\rm d}\chi^2}.
\end{equation}
At the central frequency 33.141 GHz with $\Delta_{\rm c}=-7.3\times10^{-24}$ and $\sigma_{\rm c}=3.7\times10^{-24}$, we obtained a constraint on the kinetic mixing $\chi<2.1\times10^{-12}$. The constraints on the kinetic mixing through the frequency range from 33.139 GHz to 33.143 GHz can be obtained by adopting the calculation to every frequency bin in the averaged power excess spectrum. The red line in Fig. \ref{fig:limit} shows the constraints on the kinetic mixing under the assumption of a randomly polarized scenario, where the polarization of the dark photons distribute evenly in all directions.
The results are nearly 3 orders of magnitude more strict than that provided by DPDM\cite{Arias2012WISPyCD}.

For a linear polarization scenario, the form factor is
\begin{equation}
    C=\frac{|\int dV\boldsymbol{E}|^2}{V\int dV\boldsymbol{E}^2}\langle (\boldsymbol{Z(t)}\cdot \boldsymbol{X})^2\rangle,
\end{equation}
where $\boldsymbol{Z(t)}=(\cos\lambda_{\rm lab}\cos\omega_0t, \cos\lambda_{\rm lab}\sin\omega_0t, \sin\lambda_{\rm lab})$, and $\boldsymbol{X}=(\sin\theta_{\rm DP}\cos\phi_{\rm DP}, \sin\theta_{\rm DP}\sin\phi_{\rm DP}, \cos\theta_{\rm DP})$, $\lambda_{\rm lab}$ is the latitude of our laboratory, and $\omega_0$ is the angular frequency of earth rotation\cite{PhysRevD.104.095029}. Therefore, compared to the random polarization scenario, the constraints in the linear polarization scenario will be modified by a factor $\varepsilon=\left(3\langle(\boldsymbol{Z(t)}\cdot \boldsymbol{X})^2\rangle\right)^{-1/2}$. Since the polarization direction of dark photon is unknown, we went over all the parameter space of $(\theta_{\rm DP}, \phi_{\rm DP})$ and found that the factor $\varepsilon$ for a 12 h experiment at our laboratory is in the range of (0.74, 3.26). The best constraints can be obtained if the polarization direction is nearly parallel with the z-direction of the lab frame through the experiment, as the shallow red region in Fig. \ref{fig:limit} shows.
On the contrary, the worst constraints are obtained if the polarization direction of dark photons is almost perpendicular to the z-direction of the lab frame, as the deep red region shows.

In conclusion, we report the first haloscope searching for dark photon dark matter at Q band. A 33.141 GHz cavity was designed and fabricated. Through a novel coupling mechanism separated from the cavity, tunable coupling without degrading the quality factor of the cavity was realized. In the frequency range from 33.139 to 33.143 GHz, i.e., in the dark photon mass range from 137.052 $\upmu$eV to 137.067 $\upmu$eV, the most stringent constraints on the kinetic mixing have been established. Our results surpass the previous constraints by nearly three orders of magnitude. The constraints can be further improved via microwave amplifiers with lower noise temperature and cavities with larger mode volume. Our work has demonstrated the feasibility of dark matter haloscopes in the over 30 GHz region, and can be further extended to search for high-frequency gravitational waves and other well-motivated dark matter candidates, including axions and axion-like particles.

\begin{acknowledgments}
    This work was supported by NSFC (T2388102, 12205290, 12261160569), the Innovation Program for Quantum Science and Technology (2021ZD0302200), and the National Key R\&D Program of China (Grant No. 2021YFB3202800).
    X.R. thanks the support by the Major Frontier Research Project of the University of Science and Technology of China (Grant No. LS9990000002).
    Y. Y. and R. Q. K. contributed equally to this work.
\end{acknowledgments}


\end{document}